# Randomized Caching in Cooperative UAV-Enabled Fog-RAN

M. G. Khoshkholgh*, Keivan Navaie**, Halim Yanikomeroglu†, V. C. M. Leung*, and Kang G. Shin††

* Department of Electrical and Computer Engineering, the University of British Columbia, Canada
** School of Computation and Communications, Lancaster University, UK
† Department of System and Computer Engineering, Carleton University, Canada
†† Department of Electrical Engineering and Computer Science, University of Michigan, USA

*Abstract*—We consider an unmanned aerial vehicle enabled (UAV-enabled) fog-radio access network (F-RAN) in which UAVs are considered as flying remote radio heads (RRH) equipped with caching and cooperative communications capabilities. We are mainly focus on probabilistic/randomized content placement strategy, and accordingly formulate the content placement as an optimization problem. We then study the efficiency of the proposed content placement by evaluating the average system capacity and its energy-efficiency. Our results indicate that cooperative communication plays an essential role in UAV-enabled edge communications as it effectively curbs the impact of dominant Line-of-Sight (LOS) received interference. It is also seen that cooperative cache-enabled UAV F-RAN performs better in high-rise environments than dense urban and sub-urban environments. This is due to a significant reduction of the received LOS interference because of blockage by the high-rise buildings, and the performance gain of cooperative communication on the attending signal. Comparing the performances of the developed content placement strategy and conventional caching techniques shows that our proposed probabilistic/randomized caching outperforms the others in most of the practical cases.

## I. INTRODUCTION

Using low-altitude unmanned areal vehicles (UAVs), or drones, is proposed to improve the connectivity in terrestrial wireless communications [1]. Using UAV-mounted transmitters is shown in [2] to enable line-of-sight (LOS) communication to the users equipments (UEs) on the ground, thus improving the overall performance of wireless networks. UAVs can also be equipped with sophisticated processing capabilities; for instance, [3] proposed scenarios in which UAVs assist UEs by offloading computation.

The authors of [2] obtained the optimal altitude of UAVs, providing maximum coverage. 3-D placement of UAVs in an on-demand UAV-enhanced cellular network was also investigated in [4]. Using stochastic geometry, [5] studied the spectrum sharing in drone communication networks and proposed how to adjust the density of drones to preserve the required UE coverage.

Equally important is the ground-to-air (G2A) fronthaul communications between the core network/ground base stations and UAVs. Providing real-time fronthaul communication is, however, proven to be challenging and expensive [6], as currently-deployed BSs are electronically/mechanically down-tilted, and installation of new equipments is costly and time-consuming. Therefore, acting solely as a remote radio head (RRH), UAVs may not be able to realize the required seamless connectivity and high capacity for emerging services, such as on-demand video streaming, augmented and virtual reality, and on-line gaming. This becomes even more important since more than 70% of the wireless traffic is expected to be streaming services [7].

An approach to reduce the real-time reliance on the fronthaul links is to cache the popular contents at UAVs [8]. The authors of [9] used machine learning to develop resource-allocation algorithms based on the prediction of content request patterns. Cache-enabled cellular systems and device-to-device (D2D) communications have been studied widely in terrestrial systems. An optimal probabilistic/randomized caching is proposed in [10] for small-cell networks. Randomized caching was then extended to $K$-tier heterogenous cellular networks (HetNets) [11], [12]. F-RAN and cache-enabled D2D communications in HetNets were also investigated in [13].

In 5G cellular networks, F-RAN endows the networking functionalities by introducing distributed edge computing and caching at the RRHs. Reference [14] showed F-RAN to be effective in improving energy-efficiency for delivering scalable video services. The authors of [15] developed a spatial signal alignment technique in F-RAN to efficiently mitigate interference. In spite of its advantages, however, F-RAN has not yet been adopted in UAV communications within a given cooperation zone. So, we propose a UAV-enabled F-RAN augmented by caching and cooperative communications. In our system, the UAVs located in a cooperation zone will contribute in a cooperative transmission scheme to the UEs on the ground. We then develop an optimal probabilistic contents placement using stochastic geometry.

We further evaluate the efficiency of contents placement by evaluating the capacity and energy-efficiency (EE). Our results show that *i*) since UAV communication is prone to substantial line-of-sight (LOS) interference, it is crucial to adopt a large cooperation zone; *ii*) as in high-rise environments the possibility of non-LOS links increases, and the performance of caching increases in urban and sub-urban environments; *iii*) increasing altitude of UAVs often causes the reduction of performance, as it makes it easier to establish LOS communication links between farther interfering UAVs and the ground user; and *iv*) when the size of the content library is large compared to the memory size of UAVs, the developed content placement outperforms classic caching, such as most popular contents cashing, and least recently used (LRU) caching.

## II. SYSTEM MODEL

Let $\mathcal{F} = \{f_1, f_2, \ldots, f_F\}$ be the content library (catalog) where $F = |\mathcal{F}| > 0$ is the size of library, and $f_c$ is the $c$-th most popular file in the library. The content library consists of the most popular videos, web-pages, music tracks, and the like. Adopting advanced big data analytics and machine learning, the popular contents are assumed to be chosen and then sorted by their popularity in advance.

Popularity of the contents is modelled using Zipf distribution [10], [11]. Therefore, the probability that $f_m$ is requested is $a_m = m^{-\kappa}(\sum_{c=1}^{F} c^{-\kappa})^{-1}$, where $0 \leq \kappa \leq 2$ is the skewness of the distribution referred to as *popularity exponent*. For $\kappa \to 0$, all contents become equally popular, i.e., uniform distribution. For simplicity, as in [10], [16], we assume the equal size files in $\mathcal{F}$.

We use stochastic geometry as a tool for analyzing UAV networks. The location of a UAV in the 3-D space is modeled as a Homogenous Poisson Point Process (HPPP). Let $\Phi = \{(X_i, H) \in \mathbb{R}^3, i = 1, 2, \ldots\}$, where $X_i \in \mathbb{R}^2$ is the location of UAV $i$ is the 2-D plane and $H$ is its altitude. We further assume that UAVs are all at the same constant altitude. As shown in [5], it is straightforward to extend this model to the cases of UAVs being at random altitudes. The density of UAVs is $\lambda$ per km$^2$.

The channel between the UAVs and the users on the ground, referred to as the *A2G channel*, is modeled as a combination of a large-scale path-loss attenuation, a large-scale shadowing, and a small-scale fading component [6], [4]. The A2G channel operates in LOS/NLOS modes [4], and the occurrence of LOS mode is shown to be dependent, among other things, on the drone's height, elevation angle, and environment, e.g., dense urban or sparse rural. The probability that the channel between UAV $X_i$ and a receiver located at the origin, referred to as a *typical UE*, is an LOS channel specified by a distance-dependent probability:

$$p_L(\|X_i\|) = \left(1 + \phi e^{-\psi\left(\frac{180}{\pi}\arctan(\frac{H}{\|X_i\|}) - \phi\right)}\right)^{-1}, \quad (1)$$

where $\|X_i\|$ is the 2-D Euclidian distance between the typical user and the drone, and $\phi$ and $\psi$ are the channel parameters capturing the traits of the underlying communication environment. Increasing $H$ increases the probability of experiencing an LOS channel. Using (1), the path-loss attenuation is:

$$L(\|X_i\|) =$$
$$\begin{cases} L_L(\|X_i\|) = K_L(\sqrt{H^2 + \|X_i\|})^{-\alpha_L} & \sim p_L(\|X_i\|), \\ L_N(\|X_i\|) = K_N(\sqrt{H^2 + \|X_i\|})^{-\alpha_N} & \sim p_N(\|X_i\|), \end{cases}$$
(2)

where $\alpha_L$ ($\alpha_N$) is the LOS (NLOS) path-loss exponent and $K_L$ ($K_N$) is the corresponding intercept constant. Note that increasing $H$ results in a higher attenuation as the signal needs to traverse further, thus losing its power and causing the UAV to consume more energy. At the same time, a larger $H$ is advantageous as it may make the LOS component dominant where its path-loss exponent is much smaller than that of the NOLS. In fact, balancing the required transmit power on one hand and the channel attenuation advantage on the other hand, leads to an optimal $H$; see, e.g., [2], [4].

TABLE I
AIR-TO-GROUND PARAMETERS AND THE CORRESPONDING VALUES [2].

|  | High-Rise | Dense-Urban | Urban | Sub-Urban |
|---|---|---|---|---|
| $\phi$ | 27.23 | 12.08 | 9.61 | 4.88 |
| $\psi$ | 0.08 | 0.11 | 0.16 | 0.43 |
| $\mu_L$ | 1.5 | 1 | 0.6 | 0 |
| $\mu_N$ | 29 | 20 | 17 | 18 |
| $a_L$ | 7.37 | 8.96 | 10.39 | 11.25 |
| $a_N$ | 37.08 | 35.97 | 29.6 | 32.17 |
| $c_L$ | 0.03 | 0.04 | 0.05 | 0.06 |
| $c_N$ | 0.03 | 0.04 | 0.03 | 0.03 |

For the fading fluctuations, we consider both small-scale power fading, $W_X$, and large-scale shadowing, $V_X$. The former is modeled using Nakagami fading:

$$W_X = \begin{cases} W_X^L = \Gamma(\overline{W}_L, \frac{1}{\overline{W}_L}) & \sim p_L(X) \\ W_X^N = \Gamma(\overline{W}_N, \frac{1}{\overline{W}_N}) & \sim p_N(X), \end{cases} \quad (3)$$

where $\Gamma(a, b)$ is the Gamma distribution with parameters $a$ and $b$. Depending on the LOS/NLOS status of the communication channel between the UAV $X$ and the UE, the parameters $a, b$ will be different. It is reasonable to assume that $\overline{W}_L > \overline{W}_N$ as the fading is often more severe in NLOS channels.

For the large-scale shadow-fading, we adopt a Log-normal model with the shadowing power gain $V_X = 10^{U_X}$:

$$U_X = \begin{cases} U_X^L \sim \mathcal{N}(\mu^L, \sigma_X^L) & \sim p_L(X) \\ U_X^N \sim \mathcal{N}(\mu^N, \sigma_X^N) & \sim p_N(X), \end{cases} \quad (4)$$

and $\mathcal{N}(\mu, \sigma)$ denotes a normal distribution with mean $\mu$ and variance $\sigma^2$, and $\sigma_X^{n_X}$ is given in [2]:

$$\sigma_X^{n_X} = a_{n_X} e^{-c_{n_X} \frac{180}{\pi} \arctan(\frac{H}{\|X\|})}, \quad (5)$$

where $n_X = L$ (resp. $N$) represents LOS (NLOS) channels and $a_{n_X}$ and $c_{n_X}$ are channel parameters that depend on the communication environment. Table I provides $a_{n_X}$ and $c_{n_X}$ for several A2G communication scenarios.

## III. OPTIMAL PROBABILISTIC PLACEMENT OF CONTENTS

The operation of UAV-enabled F-RAN is divided in two phases: *content placement*, and *content delivery*. The content placement phase commonly operates during off-the-peak traffic periods. In this phase, the popular contents are cached in the UAVs — we assume uncoded caching so as to fully cache the files, and the UAV cache size is $S$. In the content delivery phase, the network delivers the requested files to the UEs on the ground using cooperative communications.

We consider a cooperative communication scenario in which the requested content, $f_c$, is delivered to the UE by several adjacent UAVs, where distributed beamforming is adopted and close-enough UAVs simultaneously transmit the same content to the UE. A UAV $X_i$ with circular cooperative zone of $X_{cop}$

(km) transmits $f_c$ to the typical user located at the origin, if $f_c$ exists in its cache, and $\|X_i\| \leq X_{cop}$.

For cache-enabled terrestrial communications, various techniques have been proposed, see, e.g., [10], [11] and references therein. Here we consider randomized (probabilistic) content placement (RCP) in [10], and extend it to UAV-enabled F-RAN. In RCP, UAV $X_i$ caches $f_c$ with the probability $p_c$, where $\sum_{c=1}^{F} p_c = S$, and $S$ is the cache size. Therefore, the UAVs with $f_c$ in their cache belong to set $\Phi_c \subseteq \Phi$ and forms an HPPP with density $\lambda p_c$. Therefore, $\tilde{\Phi}_c = \{X \in \Phi_c : \|X\| \leq X_{cop}\}$, is the set of UAVs able to engage in cooperative transmission of $f_c$ to the typical user.

To place $f_c$, we need to ensure that in the formed cooperative zone, $X_{cop} > 0$, $\tilde{\Phi}_c \neq \varnothing$, thus not requiring retrieval via the backhaul. In randomized content placement, the probability that at least one UAV caches $f_c$ is $1 - e^{-\pi\lambda X_{cop}^2 p_c}$. Therefore, we obtain $p_c$s that maximize the content to be cached in the UAVs:

$$\mathcal{O}: \max_{\sum_{c=1}^{F} p_c = S, p_c \in [0,1] \forall c} \sum_{c=1}^{F} a_c \left(1 - e^{-\pi\lambda X_{cop}^2 p_c}\right).$$

Since $\mathcal{O}$, is a convex optimization problem, one can use the Lagrange method to find its solution, see for details [10].

## IV. AVERAGE SYSTEM CAPACITY

We now analyze the capacity of the optimal probabilistic content placement developed in Section III. To increase the spectral efficiency, the system utilizes coordinated multi-point (CoMP) communication in the delivery phase, where UAVs form distributed beamforming [13]. Since the content is cached at the UAVs, establishing a high-data rate backhaul links among the UAVs is not required.

For the typical ground UE located at the origin requesting $f_c$, the intended signal at the receiver is $\sum_{X \in \tilde{\Phi}_c} L(X) V_X W_X$, where $\tilde{\Phi}_c$ is the set of UAVs with $f_c$ in their cache which are also flying over the cooperation zone. Ignoring the impact of background noise (interference-limited system), the received SIR is

$$\text{SIR}_c = \frac{\sum_{X \in \tilde{\Phi}_c} L(X) V_X W_X}{\sum_{Z \in \Phi \setminus \Phi_c} L(Z) V_Z W_Z + \sum_{Y \in \Phi_c \setminus \tilde{\Phi}_c} L(Y) V_Z W_Y}. \quad (6)$$

In (6), the first term in the denominator represents the received interference from the active UAVs within the cooperation zone without $f_c$ in their cache while the second term shows the received interference from those UAVs with $f_c$ in their cache located outside of the cooperation zone.

The capacity of delivering $f_c$ is then $\overline{R}_c = \mathbf{E} \log\left(1 + \text{SIR}_c\right)$ where $\mathbf{E}$ represents expectation over any randomness including position of the UAVs, path-loss attenuation, small-scale fading, shadowing, and content placement. The total achievable system capacity is therefore $\overline{R} = \sum_{c=1}^{F} a_c \overline{R}_c$. In what follows, we derive an approximation of $\overline{R}$.

We assume that there is at least one UAV with $f_c$ in its cache, so $\overline{R}_c > 0$ and

$$\overline{R}_c = \mathbf{E}\left[1_{|\tilde{\Phi}_c| > 0} R_c\right], \quad (7)$$

where

$$R_c = \log\left(1 + \frac{\sum_{X \in \tilde{\Phi}_c} L(X) V_X W_X}{\sum_{Z \in \Phi \setminus \Phi_c} L(Z) V_Z W_Z + \sum_{Y \in \Phi_c \setminus \tilde{\Phi}_c} L(Y) V_Z W_Y}\right).$$

Using $\log(1 + x) = \int_0^\infty e^{-v}/v \left(1 - e^{-vx}\right) dv$, we write

$$R_c = \int_0^\infty \frac{e^{-v}}{v}\left(1 - e^{-v \frac{\sum_{X \in \tilde{\Phi}_c} L(X) V_X W_X}{\sum_{Z \in \Phi \setminus \Phi_c} L(Z) V_Z W_Z + \sum_{Y \in \Phi_c \setminus \tilde{\Phi}_c} L(Y) V_Z W_Y}}\right) dv$$

$$= \int_0^\infty \frac{1}{v} e^{-v \sum_{Z \in \Phi \setminus \Phi_c} L(Z) V_Z W_Z} e^{-v \sum_{Y \in \Phi_c \setminus \tilde{\Phi}_c} L(Y) V_Y W_Y} \left(1 - e^{-v \sum_{X \in \tilde{\Phi}_c} L(X) V_X W_X}\right) dv.$$

Note that $\forall m, n$, $\Phi_m$ and $\Phi_n$ are independent. $\Phi_c \setminus \tilde{\Phi}_c$ and $\tilde{\Phi}_c$ are also independent, and hence (7) is reduced to

$$\overline{R}_c = \int_0^\infty \frac{1}{v} \underbrace{\mathbf{E}_{\Phi \setminus \Phi_c, \{L(Z), V_Z, W_Z\}_{Z \in \Phi \setminus \Phi_c}}\left[e^{-\sum_{Z \in \Phi \setminus \Phi_c} vL(Z) V_Z W_Z}\right]}_{T_1}$$

$$\times \underbrace{\mathbf{E}_{\Phi_c \setminus \tilde{\Phi}_c, \{L(Z), V_Z, W_Z\}_{Z \in \Phi_c \setminus \tilde{\Phi}_c}}\left[e^{-\sum_{Y \in \Phi_c \setminus \tilde{\Phi}_c} vL(Y) V_Y W_Y}\right]}_{T_2}$$

$$\times \underbrace{\mathbf{E}_{\tilde{\Phi}_c, \{L(Z), V_Z, W_Z\}_{Z \in \tilde{\Phi}_c}}\left[1_{|\tilde{\Phi}_c| > 0}\left(1 - e^{-v \sum_{X \in \tilde{\Phi}_c} L(X) V_X W_X}\right)\right]}_{T_3} dv.$$

In the following, we evaluate $T_1$, $T_2$, and $T_3$.

For, $T_1$, we note that 1) each communication link independently undergoes LOS mode, 2) shadowing and fading power gains on each communication link are independent, and 3) shadowing (fading) power gains across communication gains are independent. Therefore, it is straightforward to show

$$T_1 = \mathbf{E}_{\Phi \setminus \Phi_c, \{L(Z), V_Z, W_Z\}_{Z \in \Phi \setminus \Phi_c}} e^{-\sum_{Z \in \Phi \setminus \Phi_c} vL(Z) V_Z W_Z}$$

$$= \mathbf{E}_{\Phi \setminus \Phi_c} \prod_{Z \in \Phi \setminus \Phi_c} \mathbf{E}_{\{L(Z), V_Z, W_Z\}} e^{-vL(Z) V_Z W_Z}$$

$$= \mathbf{E}_{\Phi \setminus \Phi_c} \prod_{Z \in \Phi \setminus \Phi_c} \left(\sum_{n_Z \in \{L, N\}} p_{n_Z}(Z) \right.$$

$$\left. \times \mathbf{E}_{V_Z^{n_Z}} \mathbf{E}_{W_Z^{n_Z}} e^{-vL_{n_Z}(Z) V_Z^{n_Z} W_Z^{n_Z}}\right)$$

$$= \mathbf{E} \prod_{Z \in \Phi \setminus \Phi_c} \left(\sum_{n_Z \in \{L, N\}} \mathbf{E}_{V_Z^{n_Z}} \frac{p_{n_Z}(Z)}{(1 + \frac{vL_{n_Z}(Z) V_Z^{n_Z}}{\overline{W}_{n_Z}})^{\overline{W}_{n_Z}}}\right)$$

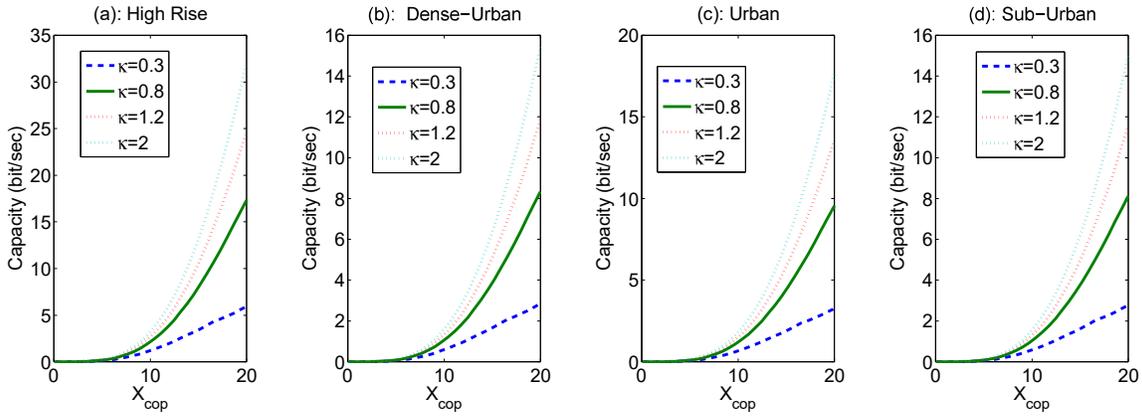

Fig. 1. Capacity (bit/sec) versus $X_{cop}$ (km), where $\lambda = 10^{-3}$, $S = 5$, and $F = 20$.

$$= \mathbf{E} \prod_{Z \in \Phi \setminus \Phi_c} \left( \sum_{n_Z \in \{L,N\}} \int_0^\infty \frac{e^{-\frac{(u-\mu^{n_Z})^2}{2(\sigma_Z^{n_Z})^2}}}{\sqrt{(2\pi\sigma_Z^{n_Z})^2}} \right.$$

$$\left. \times \frac{p_{n_Z}(Z)}{(1 + \frac{v 10^u L_{n_Z}(Z)}{\overline{W}_{n_Z}}) \overline{W}_{n_Z}} du \right)$$

$$= \exp\left\{ -2\pi(1-p_c^*)\lambda \sum_{n \in \{L,N\}} \int_0^\infty \int_0^\infty z \frac{e^{-\frac{(u-\mu^n)^2}{2(\sigma_z^n)^2}}}{\sqrt{(2\pi\sigma_z^n)^2}} \right.$$

$$\left. \times \left(1 - \frac{p_n(z)}{(1 + \frac{v L_{n_Z}(Z) 10^u}{\overline{W}_{n_Z}}) \overline{W}_n}\right) dudz \right\}, \quad (8)$$

where we also insert normal distribution with distance-dependent variance as in (5) and apply Laplace generation functional of HPPP as in [17].

Following the same line of argument, $T_2$ is also written as

$$T_2 = \exp\left\{ -2\pi p_c^* \lambda \sum_{n \in \{L,N\}} \int_{X_{cop}}^\infty \int_0^\infty z \frac{e^{-\frac{(u-\mu^n)^2}{2(\sigma_z^n)^2}}}{\sqrt{(2\pi\sigma_z^n)^2}} \right.$$

$$\left. \times \left(1 - \frac{p_n(z)}{(1 + \frac{v L_{n_Z}(Z) 10^u}{\overline{W}_{n_Z}}) \overline{W}_n}\right) dudz \right\}. \quad (9)$$

For $T_3$, in the cooperation zone, the cooperative UAVs are located randomly and the number of them is a Poisson random variable with mean $\pi \lambda X_{cop}^2 p_c^*$, and hence

$$T_3 = \mathbf{E}_{\tilde{\Phi}_c, \{L(Z), V_Z, W_Z\}_{Z \in \tilde{\Phi}_c}} \left[ 1_{|\tilde{\Phi}_c| > 0} \left(1 - e^{-v \sum_{X \in \tilde{\Phi}_c} L(X) V_X W_X}\right) \right]$$

Unlike $T_1$, and $T_2$, it is not easy to obtain the above expression numerically. We propose the following approximation to deal with this difficulty:

$$T_3 \approx \mathbf{P}\{|\tilde{\Phi}_c| > 0\} \mathbf{E}_{\tilde{\Phi}_c, \{L(Z), V_Z, W_Z\}_{Z \in \tilde{\Phi}_c}} \left[1 - e^{-v \sum_{X \in \tilde{\Phi}_c} L(X) V_X W_X}\right]$$

$$= (1 - e^{-\pi \lambda X_{cop}^2 p_c^*}) \left(1 - \exp\left\{-2\pi p_c^* \lambda \sum_{n \in \{L,N\}} \int_0^{X_{cop}} \int_0^\infty z \right.\right.$$

$$\times \frac{e^{-\frac{(u-\mu^n)^2}{2(\sigma_z^n)^2}}}{\sqrt{(2\pi\sigma_z^n)^2}} \left(1 - \frac{p_n(z)}{(1 + \frac{v L_{n_Z}(Z) 10^u}{\overline{W}_{n_Z}}) \overline{W}_n}\right) dudz \right\} \bigg). \quad (10)$$

## V. ENERGY-EFFICIENCY

In UAV-enabled communications, energy-efficiency is equally important as drones have a limited energy storage [1], [8]. We also investigate the energy-efficiency (EE) of the edge caching under consideration.

To deliver $f_c$, the transmission and circuit power of all UAVs in the cooperation zone should be considered in the formulation of EE. Caching also consumes energy, where the amount of the required energy depends on the cache size and its underlying memory technology, e.g., solid state disk (SSD), dynamic random access memory (DRAM) [18].

Let the required power by a unit of cache be $P_p$ Watts. Then, a UAV consumes $SP_p$ for caching. There are also static and dynamic circuit power consumptions, where the former, denoted by $P_s$, is often constant, and the latter, denoted by $\overline{R}_c$, is related to the UAV's transmission rate. It is shown in [19], [20] that $\overline{R}_c = \Delta(\overline{R}_c)$, where $\Delta(.)$ is an increasing function of $\overline{R}_c$. We consider a linear model where $\Delta(\overline{R}_c) = \zeta \overline{R}_c$, and $\zeta$ is a system parameter. The system EE is then formulated as

$$\eta = \sum_{c=1}^F a_c \sum_{k=1}^\infty \frac{1}{k!} \frac{(\pi \lambda (X_{cop})^2 p_c^*)^k e^{-\pi \lambda (X_{cop})^2 p_c^* \overline{R}_c}}{k(P + SP_p + P_s) + \Delta(\overline{R}_c)} \quad (11)$$

which is a function of cache size $S$, altitude $H$, and the size of cooperation zone. By incorporating the approximation for $\overline{R}_c$ in Section IV, EE is then approximated as

$$\eta \approx \sum_{c=1}^F a_c \sum_{k=1}^\infty \frac{(\pi \lambda X_{cop}^2 p_c^*)^k}{k!} \frac{e^{-\pi \lambda X_{cop}^2 p_c^* \overline{R}_c}}{\frac{k(P+SP_p+P_s)}{(1-e^{-\pi \lambda X_{cop}^2 p_c^*})} + \zeta \overline{R}_c}. \quad (12)$$

## VI. NUMERICAL RESULTS AND COMPARISONS

We now evaluate the impact of various system parameters on the approximated achievable capacity and EE. The parameters of the A2G channel are listed in Table I. We also set $\alpha_L = 2.09$, $\alpha_N = 4$, $\overline{W}_L = 10$ and $\overline{W}_N = 2$. A simple channel access mechanism is also considered where the available radio spectrum is divided into $B = 64$ sub-channels and UAVs randomly choose a sub-channel for transmission.

Fig. 1 shows the impact of cooperation zone radius on the average system capacity for different values of popularity exponent, $\kappa$, in several communication environments. As shown in Fig. 1, increasing $X_{cop}$ results in a higher average capacity as it improves the attending signal strength over the interference, and also improves the chance of the requested content being cached at the UAVs in the cooperation zone. Since both attending and interference signals are likely to experience LOS channels, for a smaller $X_{cop}$, the UE may receive interfere signals from a larger number of the UAVs. Therefore, a large cooperation zone is needed to reduce the impact of interference and improve the signal strength through cooperation. Fig. 1 further shows the impact of popularity exponent; increasing $\kappa$ results in a higher average system capacity. For a smaller $X_{cop}$, the average system capacity is almost 0 for any $\kappa$. For $\kappa > 0.3$, Fig. 1 also shows the same trend in different environments:

$$\frac{\overline{R}_{\text{High-Rise}}}{2} \gtrapprox \overline{R}_{\text{Dense-Urban}} \approx \overline{R}_{\text{Urban}} \approx \overline{R}_{\text{Suburban}}.$$

This is because in high-rise building environments, the UAV signals will likely experience NLOS mode and the LOS signal could be blocked by the high-rise buildings. This can substantially reduce the received interference at the users. Higher chance of experiencing an NLOS channel is generally not problematic in this case as cooperative communication improves the strength of the attending signal.

In Fig. 2, we also study the impact of library size, $F$, on the average system capacity. Specifically, we compare the average system capacity of the proposed content placement with that of the most popular contents (MPC) and least recently used (LRU) caching. In the MPC, only the $S$ most popular contents are cached at UAVs. In the LRU, the least requested content in the cache is replaced by the newly requested content. The proposed probabilistic caching always outperforms LRU for different values of the popularity index, $\kappa$. In general, we observe that by increasing $F$ the capacity increases regardless of $\kappa$ when the contents are placed randomly. Nevertheless, for the MPC growing $F$ render smaller capacity. Last, under LRU the capacity may grow/decline by $F$ depending on the value of $\kappa$.

On the other hand, we observe that probabilistic algorithm always outperforms LRU, where for sufficiently large values of $F$ the maximum gain is actually measured. Comparing to MPC, we see that when $F$ is sufficiently small, the randomized scheme has slightly smaller capacity. However for library sizes larger than $F > 3S$ the latter always outperforms the former.

Using (12) we investigate the impact of system and design parameters on EE. Fig. 3 plots EE vs. the radius of cooperation zone for different content popularity exponents in several communication environments. Fig. 3 suggests that increasing the size of cooperation zone leads to a higher EE. We also observe that in a high-rise communication environment, cooperative caching results in a much higher EE than in other environments.

Fig. 4 shows the impact of UAVs' altitude, $H$, on the EE. As expected, our results confirm that increasing $H$ results in a lower EE due mainly due to the reduction of capacity; as the higher the altitude, the larger is the path-loss attenuation. The results in Fig. 4 further suggest that in high-rise environments, the rate of EE reduction due to increasing $H$ is much lower than in other environments. This is mainly because in such environments, the received interference is likely to be weaker as the signals from some of the interfering UAVs might be blocked by tall buildings. The attending signal will be also weaker, but this is compensated through the cooperative communication.

## VII. CONCLUSIONS AND DISCUSSIONS

We explored a cache-enabled UAV network with F-RAN technology serving users on the ground. We considered a cooperative communications and investigated placement and delivery of contents. We used stochastic geometry to derive the optimal content placement probability. We studied the efficiency of content placement by evaluating the average system capacity and energy-efficiency. Our results indicate that cooperative communication plays an essential role in UAV-enabled edge communications.

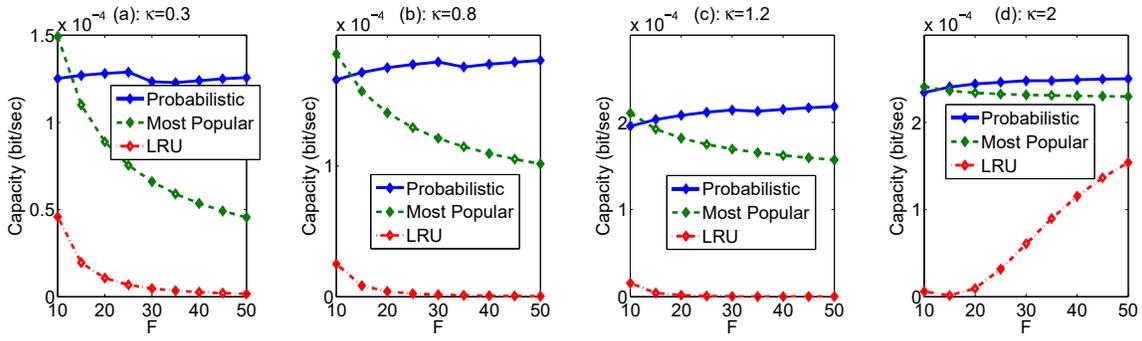

Fig. 2. Average system capacity versus $S/F$ for $S = 5$, $X_{cop} = 1$ km, $\lambda = 10^{-3}$ in sub-urban area.

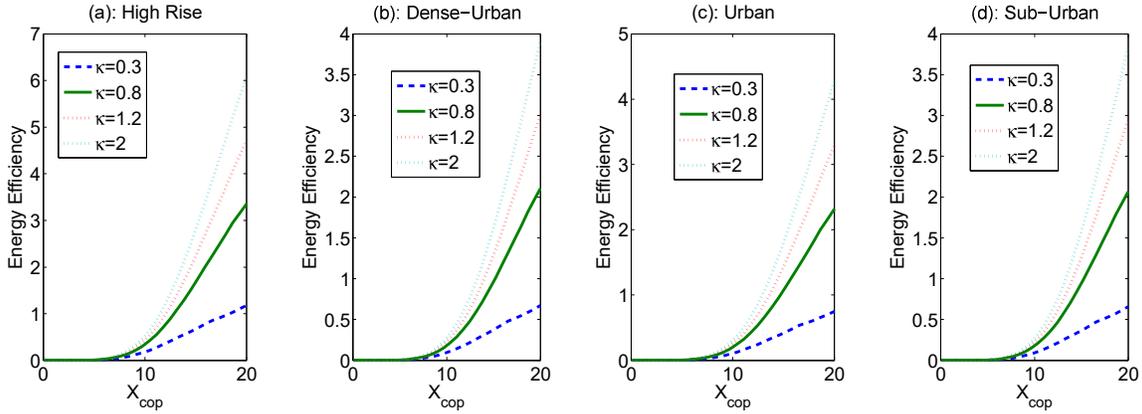

Fig. 3. Energy-efficiency versus $X_{cop}$, where $\lambda = 10^{-3}$, $S = 5$, and $F = 20$.

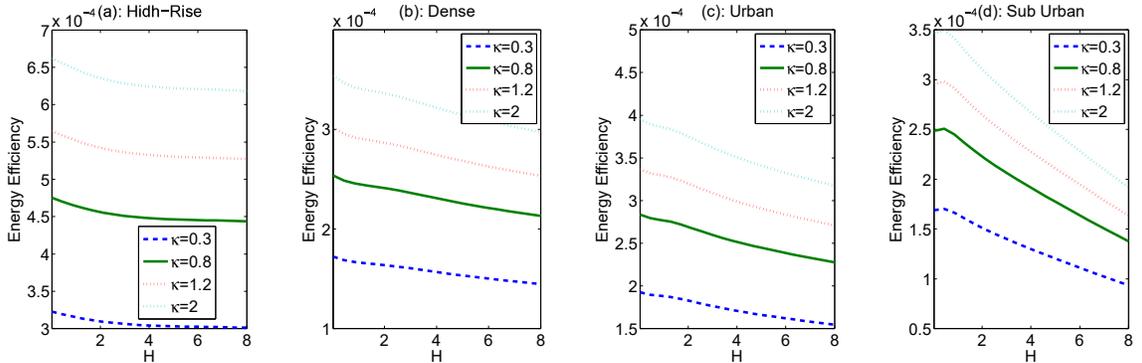

Fig. 4. Energy-efficiency versus the altitude of UAVs, $H$ (km), where $X_{cop} = 3$km, $\lambda = 10^{-3}$, $S = 5$, and $F = 20$.